\documentclass[aps,preprint,pra]{revtex4-1}
\usepackage{graphicx}
\usepackage{amssymb}
\usepackage{amsmath}
\usepackage{color}
\usepackage{enumerate}
\usepackage{bm}
\usepackage[utf8]{inputenc}
\begin{document}
\title{ Anisotropic Effect of Dipolar Interaction in Ordered Ensembles of Nanoparticles}
\author{Manish Anand}
\email{itsanand121@gmail.com}
\affiliation{Department of Physics, Bihar National College, Patna University, Patna-800004, India.}

\date{\today}

\begin{abstract}
We implement extensive computer simulations to investigate the hysteresis characteristics in the ordered arrays ($l^{}_x\times l^{}_y$) of magnetic nanoparticles as a function of aspect ratio $A^{}_r=l^{}_y/l^{}_x$, dipolar interaction strength $h^{}_d$, and external magnetic field directions. We have considered the aligned anisotropy case, $\alpha$ is the orientational angle. It provides an elegant en route to unearth the explicit role of anisotropy and dipolar interaction on the hysteresis response in such a versatile system. The superparamagnetic character is dominant with weak dipolar interaction ($h^{}_d\leq0.2$), resulting in the minimal hysteresis loop area. Remarkably, the double-loop hysteresis emerges even with moderate interaction strength ($h^{}_d\approx0.4$), reminiscent of antiferromagnetic coupling. These features are strongly dependent on $\alpha$ and $A^{}_r$. Interestingly, the hysteresis loop area increases with $h^{}_d$, provided $A^{}_r$ is enormous, and the external magnetic field is along the $y$-direction. The coercive field $\mu^{}_oH^{}_c$, remanent magnetization $M^{}_r$, and the heat dissipation $E^{}_H$ also depend strongly on these parameters. Irrespective of the external field direction and weak dipolar interaction ($h^{}_d\leq0.4$), there is an increase in $\mu^{}_oH^{}_c$ with $h^{}_d$ for a fixed $\alpha$ and $A^{}_r\leq4.0$. The dipolar interaction also elevates $M^{}_r$ as long as $A^{}_r$ is huge and the field is along the $y$-direction. $E^{}_H$ is minimal for negligible and weak dipolar interaction, irrespective of $A^{}_r$, $\alpha$, and the field directions. Notably, the magnetic interaction enhances $E^{}_H$ if $A^{}_r$ is enormous and the magnetic field is along the long axis of the system. These results are beneficial in various applications of interest such as digital data storage, spintronics, etc.
\end{abstract}
\maketitle
\section{Introduction}
Ordered arrays of magnetic nanoparticles (MNPs) have received significant attention due to their intriguing physics and numerous technological applications~\cite{bigioni2006,jiang2014,haakonsen2021,polarz2011,sacanna2013}. Such systems have unique physical and chemical properties that are utterly different from their corresponding bulk counterparts~\cite{boker2007,horechyy2010}. They are also of immense importance in various applications such as spintronics, magnetic hyperthermia, drug delivery, biomedicine,  etc.~\cite{gandhi2018,reiss2005,couvreur2013,anand2016,pankhurst2003}. Therefore, understanding of various magnetic properties of these nanosystems is the need of time.

The physics of non-interacting nanoparticle arrays are fairly known~\cite{carrey2011}. However, MNPs interact because of the dipolar interaction in such a system~\cite{poddar2002}.  The dipolar interaction affects the systematic properties significantly in such an ordered system due to the long-range and anisotropic nature~\cite{anand2021thermal,ovejero2016}. For instance, it induces spin-glass like characteristics in MNPs assembly with positional defects~\cite{winkler2008,morup1994}. On the other hand, it promotes ferromagnetic coupling in highly anisotropic systems, such as the linear and columnar arrangement of MNPs~\cite{anand2020}. The dipolar interaction strongly affects the magnetic characteristics in a dense assembly~\cite{kostopoulou2014,torche2020,masunaga2011,kechrakos2008,figueiredo2007,myrovali2016}. Torche {\it et al.} studied the local heat dissipation in a dipolar interacting nanoparticle assembly using transition state theory~\cite{torche2020}. The dipolar interaction is found to enhances heat dissipation. Masunaga {\it et al.} investigated the dipolar interaction effect on the magnetic properties in Nickel nanoparticles ensembles~\cite{masunaga2011}. The hysteresis characteristics such as blocking temperature are strongly affected by the dipolar interaction. Kechrakos {\it et al.} studied the magnetic and transport properties in an ordered assembly of MNPs using Monte Carlo methods~\cite{kechrakos2008}.
The dipolar interaction induces anisotropic magnetic behaviour between the in-plane and out-of-plane directions of the system.  Figueiredo {\it et al.} found an enhancement in the blocking temperature because of dipolar interaction~\cite{figueiredo2007}. Myrovali {\it et al.} observed an elevation in the area under the hysteresis curve with dipolar interaction~\cite{myrovali2016}. The dipolar interaction also dictates the ground state spin morphologies~\cite{de1997,luttinger1946,politi2006,macisaac1996}. For example, it promotes ferromagnetic coupling in face-centred cubic while antiferromagnetic in the cubic assembly of MNPs~\cite{luttinger1946}. The most favourable configuration is ferromagnetic in the triangular lattice~\cite{politi2006}. In contrast, the minimum energy state is antiferromagnetic in the square array of nanoparticles~\cite{macisaac1996}.



Moreover, the dipolar interaction induces the anisotropic properties by creating an additional anisotropy termed as shape anisotropy in such a system~\cite{schmool2007,martinez2013,anand2021hysteresis}. Therefore, these ordered nanoparticles ensembles provide a rich theoretical framework to study the role of dipolar interaction and magnetic anisotropy on magnetic response. Much research efforts ranging from experimental to theoretical modelling have been devoted to unearthing the magnetic properties in such versatile systems~\cite{deng2020,li2020,wen2017,allia2020,dyab2009,hoffelner2015,anand2021relaxation}. Deng {\it et al.} presented a detailed and very informative overview regarding forming an anisotropic assembly of MNPs~\cite{deng2020}. The dipolar interaction is one of the main reasons for such a self-assembled system. Li {\it et al.} analyzed the magnetic properties in dense arrays of magnetic nanoparticles using experiments~\cite{li2020}. They are found to possess enhanced magnetic properties along the array direction compared with the dispersed particles. Wen {\it et al.} devised procedures to manipulate the shape and magnetocrystalline anisotropy during the self-assembly process~\cite{wen2017}. Maximum magnetic anisotropy is observed when the easy axis of the particle is aligned along the array axis of the sample. Allia {\it et al.} investigate the magnetic properties in an assembly of nanoparticles using the rate equation approach with collinear and  randomly distributed easy axes~\cite{allia2020}. The hysteresis loop area is significantly higher with aligned anisotropy in comparison with the randomly oriented case. Dyab {\it et al.}  fabricated the anisotropic assembly of nanoparticles.~\cite{dyab2009}. Such systems have a very high coercive field and remanent magnetization, reminiscent of ferromagnetic character. 
Hoffelner {\it et al.} investigated the orientational alignment nanoparticles using a dynamical magnetic field~\cite{hoffelner2015}. The dipolar interaction promotes the collinear arrangement of easy axes in a highly anisotropic system. In recent work, we studied the magnetic relaxation characteristics in the ordered assembly of nanoparticles using kinetic Monte Carlo simulation (kMC) with aligned anisotropy axes~\cite{anand2021relaxation}. There is a fastening or slowing down of relaxation depending on the orientation of anisotropy axes with sufficient dipolar coupling strength.


The above discussion indicates that the anisotropy axes orientations play a crucial role in determining the magnetic properties in the ordered arrays of dipolar interacting MNPs. However, a complete understanding of the effect of anisotropy axes orientations, dipolar interaction, system sizes, the direction of an external magnetic field is far from complete. Therefore, we investigate the magnetic properties in two-dimensional ordered arrays of MNPs as a function of these parameters using kinetic Monte Carlo simulations. These studies provide a deeper insight into the interplay between magnetic anisotropy and dipolar interactions on various magnetic characteristics in such a useful system. For instance, as dipolar interaction offers ferromagnetic coupling with aligned anisotropy in the columnar and linear array of magnetic nanoparticles, one can control the orientation of MNPs using an external magnetic field with ease in such cases. Such systems are advantageous in various applications such as magnetic hyperthermia, digital information storage~\cite{mehdaoui2013,krishnamurthy2008}.



The rest of the article is organized as follows. We discuss the model used and various energy terms in Sec. II.  We preset and discuss the simulation results in Sec.~III. Finally, a summary of the present work is provided in Sec. IV.

\section{Theoretical Framework}
\begin{figure}[!htb]
	\centering\includegraphics[scale=1.00]{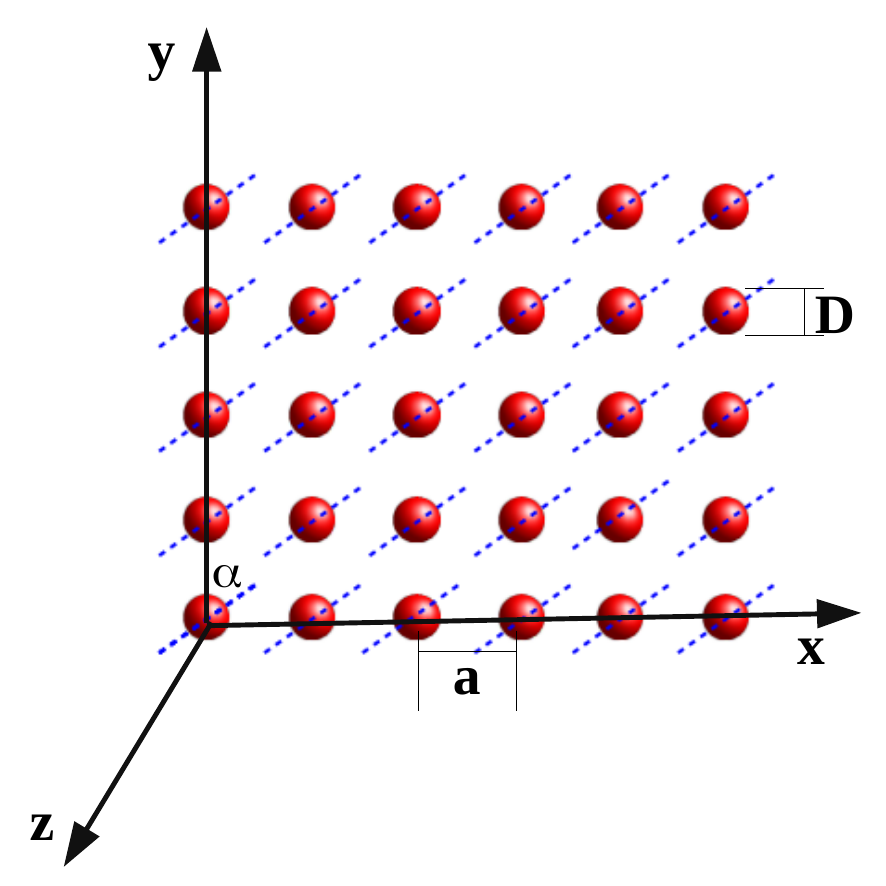}
	\caption{Schematic of the two-dimensional ordered arrays of nanoparticles. Each nanoparticle has a diameter $D$, and $a$ is the lattice constant. The blue dashed line denotes the direction of the anisotropy axis. All the MNPs are assumed to have the same anisotropy axes orientations, $\alpha$ is the orientation angle.}
	\label{figure1}
\end{figure}

We consider an ordered assembly of MNPs arranged in the two-dimensional arrays in the $xy$-plane as shown in the schematic Fig.~(\ref{figure1}). The system dimension is $l^{}_x\times l^{}_y$ and aspect ratio $A^{}_r=l^{}_y/l^{}_x$. Let the nanoparticle diameter and the lattice constant be $D$ and $a$, respectively. The particle has a magnetic moment $\mu=M^{}_sV$; saturation magnetization is $M^{}_s$, and $V=\pi D^3/6$ is the nanoparticle volume. Let the particle has magnetocrystalline anisotropy $\vec{K}=K^{}_{\mathrm {eff}}\hat{k}$, $K^{}_{\mathrm {eff}}$ is the anisotropy strength, and $\hat{k}$ is the unit vector along with the anisotropy or easy direction. We have considered the collinear anisotropy axes orientations, i.e. the anisotropy axes of all the MNPs are aligned, making an angle $\alpha$ with respect to the $y$-axis of the system [please see the schematic Fig.~(\ref{figure1})].


Each nanoparticle has the following energy because of magnetocrystalline anisotropy~\cite{carrey2011,anand2018}
\begin{equation}
E^{}_K=K^{}_{\mathrm{eff}}V\sin^2\Phi
\end{equation}
Here $\Phi$ is the angle between the anisotropy vector and the magnetic moment. As nanoparticles primarily interact because of dipolar interaction, the corresponding interaction energy can be evaluated using the following relation~\cite{kechrakos2008,usov2017}
\begin{equation}
\label{dipole}
E^{}_{\mathrm{dip}}=\frac{\mu^{}_o\pi M^{2}_sD^{6}}{144 a^3}\sum_{j,\ j\neq i}\left[ \frac{\hat{\mu_{i}}\cdot\hat{\mu_{j}}}{(r^{}_{ij}/a)^3}-\frac{3\left(\hat{\mu_{i}}\cdot\hat{r}_{ij}\right)\left(\hat{\mu_{j}}\cdot\hat{r}_{ij}\right)}{(r^{}_{ij}/a)^3}\right].
\end{equation}
Here $\mu_{o}$ is the permeability of free space; $i^{th}$ and $j^{th}$ nanoparticles have unit magnetic moment vectors $\hat{\mu}_{i}$ and $\hat{\mu}_{j}$, respectively. The centre-to-centre distance between the $i^{th}$ and $j^{th}$ magnetic moments is $r^{}_{ij}$, $\hat{r}^{}_{ij}$ is the unit vector associated with $\vec{r}_{ij}$.
In such a case, the expression of the dipolar field can be written as~\cite{tan2014}
\begin{equation}
\mu^{}_{o}\vec{H}^{}_{\mathrm {dip}}=\frac{\pi\mu_{o} M^{}_sD^{3}}{24 a^3}\sum_{j,j\neq i}\frac{3(\hat{\mu}^{}_j \cdot \hat{r}_{ij})\hat{r}^{}_{ij}-\hat{\mu^{}_j} }{(r_{ij}/a)^3}.
\label{dipolar1}
\end{equation}
We evaluate this sum precisely without Ewald summation or a cutoff radius, similar to recent works~\cite{tan2014,anand2020,anand2021hysteresis}. We define a control parameter $h^{}_d=D^{3}/a^{3}$ to model the variation of dipolar interaction strength. It correctly captures the physics of interaction strength variation as the dipolar field, and corresponding energy varies as $1/r^{3}_{ij}$ [please see  Eq.~(\ref{dipole}) and Eq.~(\ref{dipolar1})]. The particles are at the closest approach with $h^{}_d=1.0$ because $D=a$ in this case. Therefore, $h^{}_d=1.0$ captures the physics of the most substantial dipolar interacting situation. In contrast, $h^{}_d=0$ represents the non-interacting situation.


We apply an oscillating magnetic field to investigate the magnetic hysteresis behaviour in the ordered arrays of dipolar coupled MNPs. It is given by~\cite{anand2021hysteresis}
\begin{equation}
\mu^{}_{o}\vec{H}=\mu^{}_oH^{}_{\mathrm {o}}\cos(2\pi\nu t)\hat{e},
\label{magnetic}
\end{equation} 
where $\mu^{}_{o}H_{\mathrm {o}}$ and $\nu$ are the magnitude and linear frequency of the external magnetic field, respectively, and $t$ is the time. $\hat{e}$ is the unit vector along the applied oscillating magnetic field direction, which is $\hat{x}$ (along the $x$-axis) and $\hat{y}$ (along the $y$-axis) in the present work. We can then write the expression for total energy as~\cite{tan2014,anand2019}
\begin{equation}
E=K^{}_{\mathrm {eff}}V\sum_{i}\sin^2 \Phi^{}_i+\frac{\mu^{}_o\pi M^{2}_sD^{6}}{144 a^3}\sum_{j,\ j\neq i}\left[ \frac{\hat{\mu_{i}}\cdot\hat{\mu_{j}}-{3\left(\hat{\mu_{i}}\cdot\hat{r}_{ij}\right)\left(\hat{\mu_{j}}\cdot\hat{r}_{ij}\right)}}{(r_{ij}/a)^3}\right]-\mu^{}_oM^{}_sV\sum_{i}\hat{\mu}^{}_i\cdot\vec{H}
\end{equation}
Here the $i^{th}$ magnetic is inclined an angle $\Phi^{}_i$ with respect to the anisotropy axis.


We probed the hysteresis characteristics in the ordered assembly of MNPs using state of the art kinetic Monte Carlo (kMC) simulations technique. In particular, we investigate the hysteresis response as a function of aspect ratio $A^{}_r=l^{}_y/l^{}_x$, dipolar interaction strength $h^{}_d$, anisotropy orientation angle $\alpha$ and direction of the applied oscillating magnetic field. The kMC algorithm implement in the present article is described in detail in the references~\cite{tan2014,anand2021hysteresis,anand2020}. Therefore, we do not restate it to avoid duplications. It is a well-known fact that heat is dissipated due to hysteresis. The amount of heat dissipation equals the hysteresis loop area $E^{}_{H}$, which can be numerically evaluated using the following expression~~\cite{anand2016}




\begin{equation}
E^{}_{H}=\oint M(H)dH,
\label{local_heat}
\end{equation}
The above integral is calculated over one complete cycle of the external magnetic field. $M^{}(H)$ is the magnetization of system at magnetic field $H$. 

\section{Simulations Results}
We consider magnetite (Fe$^{}_3$O$^{}_4$) nanoparticles with the following values of system parameters: $D=8$ nm, $K_{\mathrm {eff}}=13\times10^3$ Jm$^{-3}$, and $M^{}_s=4.77\times10^5$ Am$^{-1}$. We have considered five values of system sizes viz $l^{}_x \times l^{}_y=20\times20$, $10\times40$, $4\times100$, $2\times200$, and $1\times400$. So, the total number of MNPs in the system is 400. The corresponding aspect ratio $A^{}_r=l^{}_y/l^{}_x$ of the system is 1.0, 4.0, 25.0, 100, and 400, respectively. The control parameter $h^{}_d$ is varied from 0 to 1.0 to model the effect of the dipolar interaction strength. The anisotropy axis orientation angle $\alpha$ is varied between 0 to $90^\circ$. Therefore, it captures the physics of perfectly aligned to perpendicularly aligned anisotropy axes. All the simulations are performed at temperature $T=300$ K. We have applied the oscillating magnetic field along $x$ and $y$-direction with respect to the sample. The magnetic field strength $\mu^{}_oH^{}_{\mathrm {o}}$ and $\nu$ is taken as $0.10$ T and $10^5$ Hz, respectively.

\begin{figure}[!htb]
\centering\includegraphics[scale=0.50]{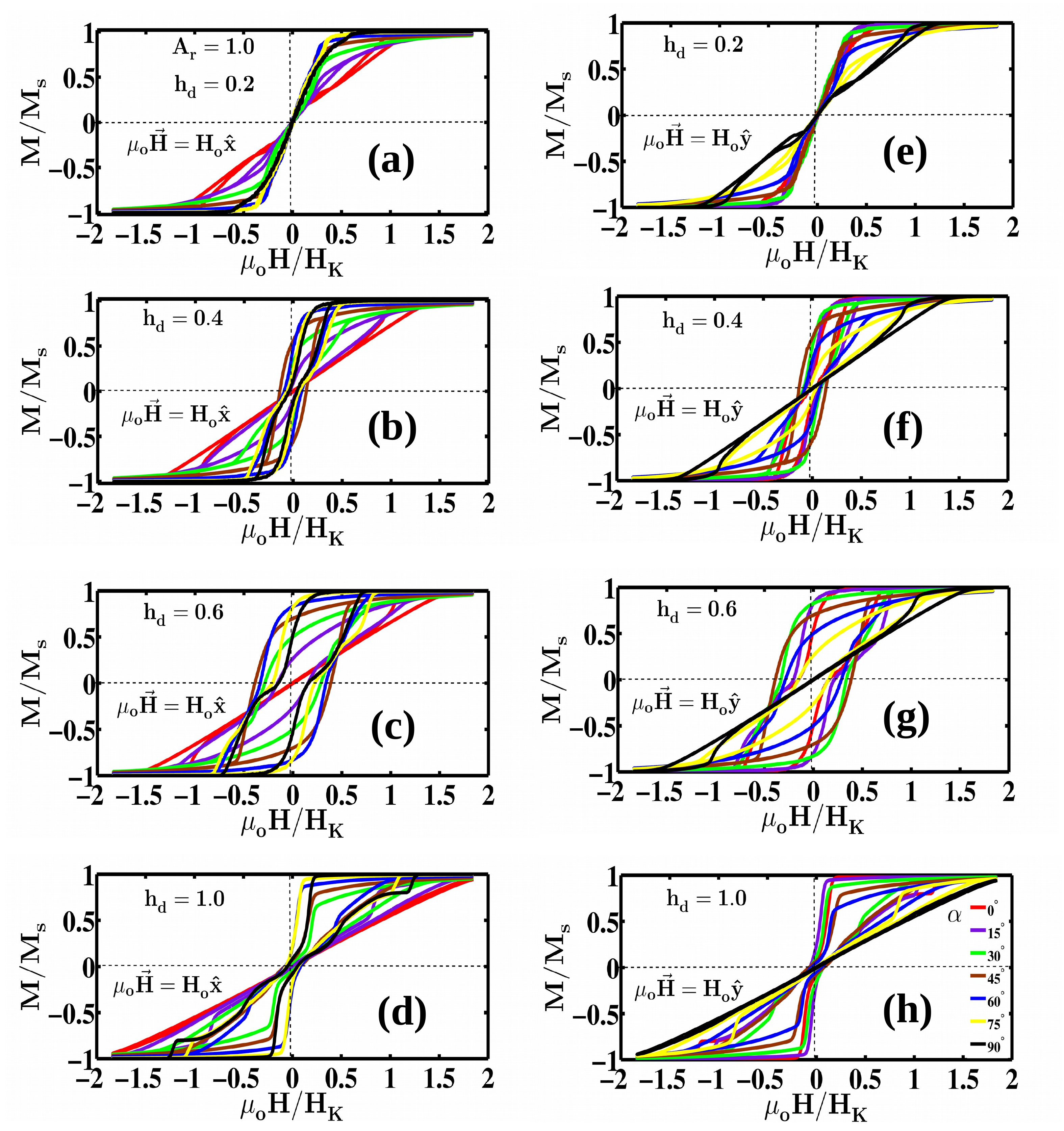}
\caption{Magnetic hysteresis in the square array of nanoparticles as a function of dipolar interaction strength and external field direction. We have taken into account four values of interaction strength: $h^{}_d=0.2$ [(a) and (e)], $h^{}_d=0.4$ [(b) and (f)], $h^{}_d=0.6$ [(c) and (g)], and $h^{}_d=1.0$ [(d) and (h)]. Double loop hysteresis emerges even with moderate $h^{}_d=0.4$, signature of antiferromagnetic coupling.}
\label{figure2}
\end{figure}

First, we investigate the hysteresis behaviour in the square array of nanoparticles as a function of dipolar interaction strength, anisotropy angle and direction of the external magnetic field. Fig.~(\ref{figure2}) shows the magnetic hysteresis curve as a function of $\alpha$ with four representative values of $h^{}_d=0.2$, 0.4, 0.6 and 1.0. The alternating magnetic field is applied along $x$ and $y$-direction. The magnetic field ($x$-axis) and magnetization axes ($y$-axis) are rescaled by the single-particle anisotropy field $H^{}_K=2K_{\mathrm {eff}}/M^{}_s$ and $M^{}_s$, respectively. We can infer the following observations from hysteresis curves with the magnetic field along $x$-direction: (1) The hysteresis loop area decreases as $\alpha$ is varied from 0 to $90^\circ$ with weak dipolar interaction ($h^{}_d\leq0.2$). (2) The double-loop hysteresis curve emerges with aligned anisotropy ($\alpha=0^\circ$) and moderate interaction strength $h^{}_d=0.4$, which is an indication of the antiferromagnetic coupling dominance. (3) The hysteresis loop area also increases with $\alpha$ for substantial $h^{}_d$. (4) The hysteresis curve has all the signatures of antiferromagnetic coupling dominance with the most substantial dipolar interaction strength ($h^{}_d=1.0$), irrespective of $\alpha$. Remarkably, the hysteresis behaviour with the field applied along $y$-direction and $\alpha=0^\circ$ is precisely similar to that of $\alpha=90^\circ$ and $\mu_o\vec{H}=H^{}_o\hat{x}$, irrespective of $h^{}_d$. Similar observations can also be made for other complementary angles $\alpha$. The emergence of double loop hysteresis is a peculiar characteristic of antiferromagnetic interaction. It can be explained by the fact that dipolar interaction induces a biasing field in the plane of the ordered arrays. In contrast, the anisotropy field instigates the magnetization to align along a particular direction depending on the orientational angle $\alpha$. Consequently, the magnetization follows the external magnetic field as long as its strength is less than the coercive field, i.e. $\mu^{}_oH/H^{}_K< 0.5$. The dipolar interaction dominates the hysteresis for $\mu^{}_oH/H^{}_K> 0.5$, resulting in a spontaneous transition in magnetization orientation. Yang {\it et al.} also observed similar double loop hysteresis in the  thin-film~\cite{yang2002}. Our observations are in excellent agreement with them. The observation of antiferromagnetic coupling dominance in the square arrangement of MNPs is also in excellent agreement with the work of Ewerlin {\it et al.} and Chen {\it et al.}~\cite{ewerlin2013,chen2017}.

\begin{figure}[!htb]
\centering\includegraphics[scale=0.50]{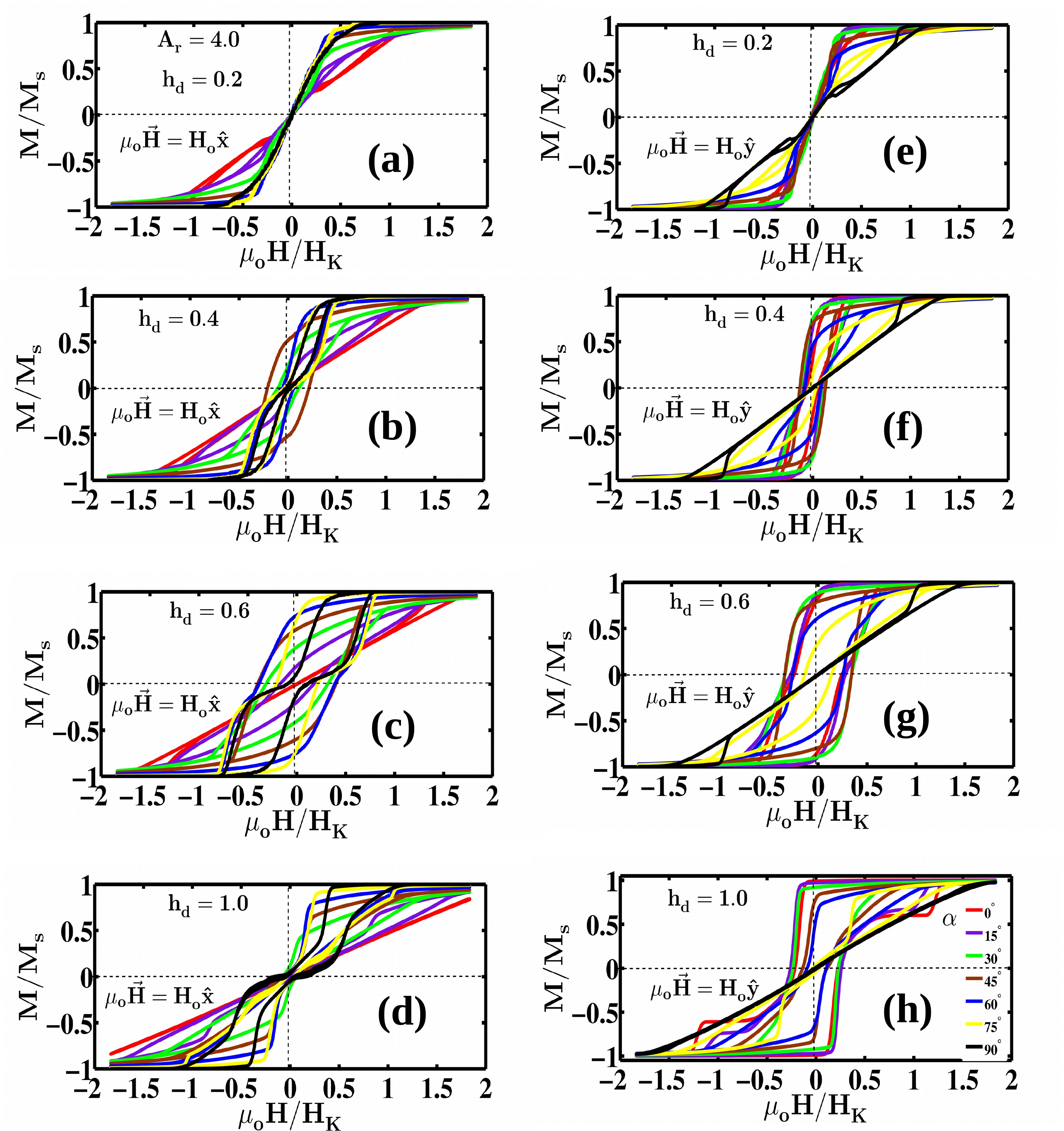}
\caption{Hysteresis response in a system with aspect ratio $A^{}_r=4.0$ as a function of $h^{}_d$ and external field direction. We have taken four values of interaction strength: $h^{}_d=0.2$ [(a) and (e)], $h^{}_d=0.4$ [(b) and (f)], $h^{}_d=0.6$ [(c) and (g)], and $h^{}_d=1.0$ [(d) and (h)]. The hysteresis loop area is minimal with small $h^{}_d$, independent of the external field direction. The antiferromagnetic coupling is dominant with appreciable $h^{}_d$, resulting in double-loop hysteresis.}
\label{figure3}
\end{figure}

The anisotropy axis direction should affect the hysteresis mechanism in the rectangular array of MNPs. So, we now probe the hysteresis response in a rectangular arrangement of nanoparticles. We plot the hysteresis curve with aspect ratio $A^{}_r=4.0$ in Fig.~(\ref{figure3}). We have used same set of other parameters 
as that of Fig.~(\ref{figure2}). Irrespective of the external magnetic field direction, the hysteresis loop area decreases with $\alpha$ for small $h^{}_{d}$. Interestingly, the hysteresis curve's shape changes from a line to the double-loop hysteresis as $\alpha$ and $h^{}_d$ are increased with the external alternating magnetic field along the $x$-axis (shorter length) of the system. Remarkably, the hysteresis loop area increases with $h^{}_d$ and collinear anisotropy axis ($\alpha=0^\circ$), provided the magnetic field is applied along the $y$-direction (long axis of the system). It is because ferromagnetic coupling gets enhanced with $h^{}_d$ in such an anisotropic system. Interestingly, double loop hysteresis tends to emerge even with moderate $h^{}_d=0.4$ as $\alpha$ is varied from 0 to $90^\circ$. In such a case, the nature of the dipolar interaction changes from ferromagnetic to antiferromagnetic as the orientation of the anisotropy axis varied from perfectly collinear to perpendicularly aligned~\cite{anand2019}. Our results with $\alpha=0^\circ$ and $90^\circ$ are in perfect agreement with the work of Li {\it et al.} and Yoshida {\it et al.}~\cite{li2020,yoshida2017}. We could not compare for other $\alpha$ as they have shown results only with two extremum $\alpha$ values. These observations also agree well with the work of Sahoo {\it et al.}~\cite{sahoo2004}.

\begin{figure}[!htb]
\centering\includegraphics[scale=0.50]{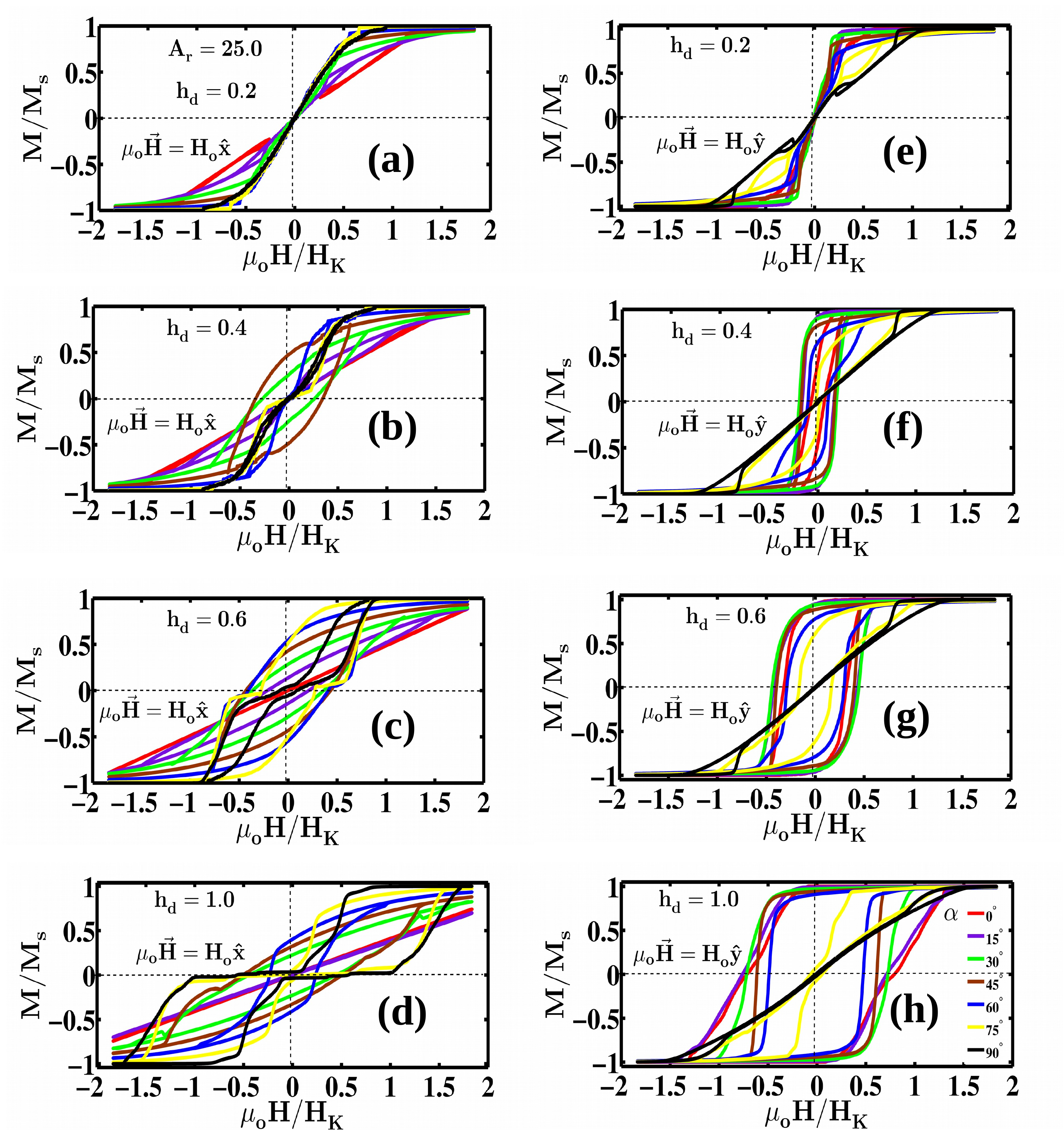}
\caption{Hysteresis curves with aspect ratio $A^{}_r=25.0$ as a function of magnetic interaction strength $h^{}_d$ and external field direction. We have taken four values of interaction strength: $h^{}_d=0.2$ [(a) and (e)], $h^{}_d=0.4$ [(b) and (f)], $h^{}_d=0.6$ [(c) and (g)], and $h^{}_d=1.0$ [(d) and (h)]. The double loop hysteresis emerges for appreciable $h^{}_d$ with external field along $x$-direction. The ferromagnetic coupling starts to dominant the hysteresis with external field applied along the $y$-axis.}
\label{figure4}
\end{figure}

\begin{figure}[!htb]
\centering\includegraphics[scale=0.40]{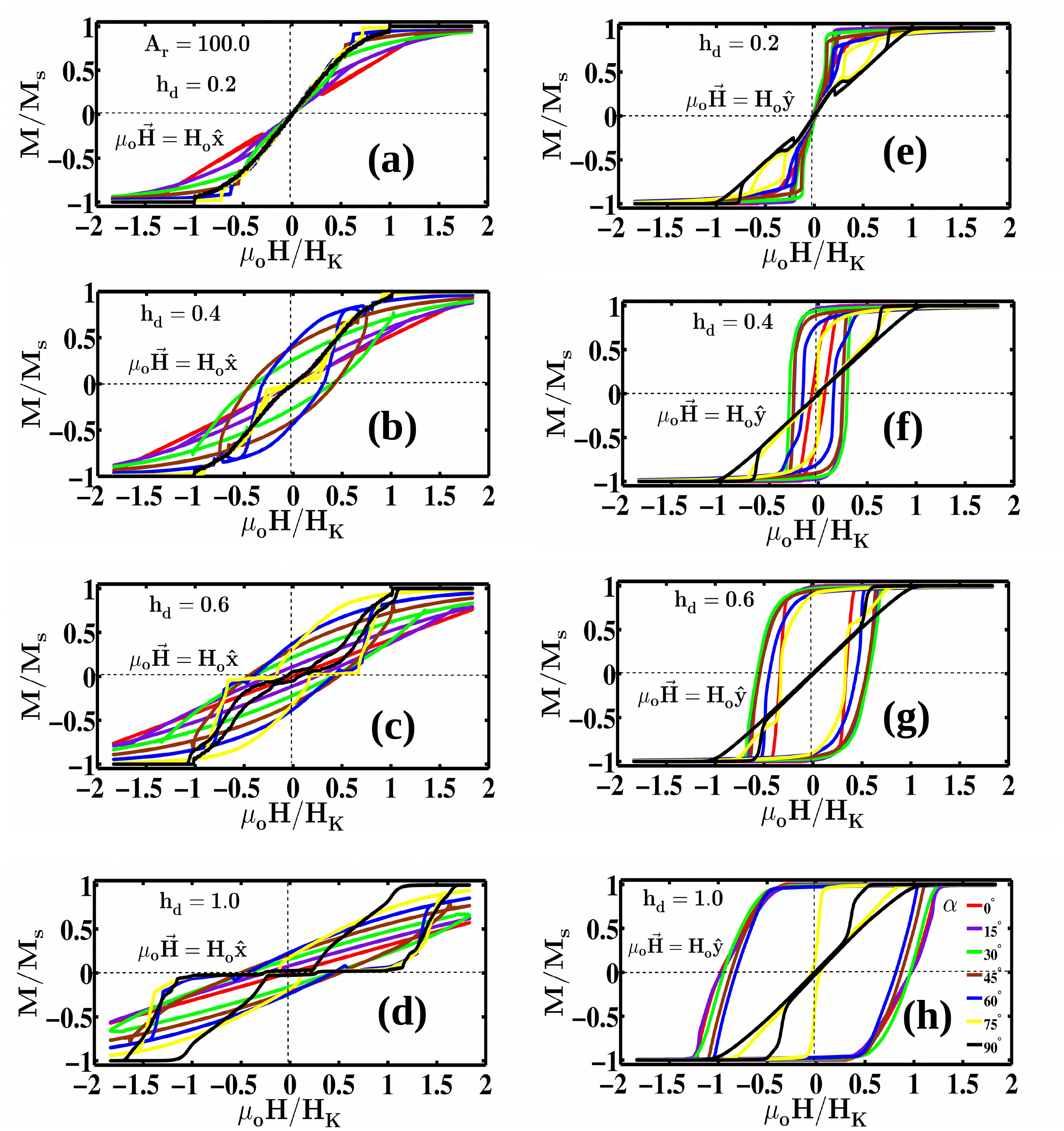}
\caption{Hysteresis curves in a system with very large aspect ratio $A^{}_r=100.0$ as a function of $h^{}_d$ and external field direction. We have taken into account four values of interaction strength: $h^{}_d=0.2$ [(a) and (e)], $h^{}_d=0.4$ [(b) and (f)], $h^{}_d=0.6$ [(c) and (g)], and $h^{}_d=1.0$ [(d) and (h)]. Non-hysteresis is observed with external field along $x$-axis and $\alpha\leq45^\circ$. The dipolar interaction induces ferromagnetic interaction with external field along the $y$-direction and $\alpha<90^\circ$.}
\label{figure5}
\end{figure}

Next, we study the magnetic hysteresis behaviour in systems with very large aspect ratios. We plot the hysteresis curves with $A^{}_r=25.0$ and 100.0 in Fig.~(\ref{figure4}) and Fig.~(\ref{figure5}), respectively. 
One can draw the following vital conclusions from these curves: (i) In the case of aligned anisotropy ($\alpha=0^\circ$) and the external magnetic field along the $x$-direction, the magnetization ceases to follow the external field, resulting in the non-hysteresis behaviour. (ii) The shape of the hysteresis curve also changes from a straight line to double loop hysteresis with $\alpha$ and $h^{}_d$ in such case. (iii) The hysteresis loop area also increases with $h^{}_d$ and aligned anisotropy axis ($\alpha=0^\circ$), provided the external field is along the $y$-direction. (iv) There is also a decrease in the hysteresis loop area as $\alpha$ is varied from $0^\circ$ to $90^\circ$. Moreover, the dominance of antiferromagnetic coupling increases with $\alpha$ and substantial $h^{}_d$. Consequently, the double hysteresis loop emerges even with moderate interaction strength ($h^{}_d\ge0.4$). In  recent work, Yuan {\it et al.} fabricated similar thin films of nanoparticles~\cite{yuan2017}. Similar MNPs ensembles are also found in biosystems, such as in the brain of migratory birds~\cite{yuan2017}. Our results could be extremely useful in assessing the hysteresis characteristics of such an assembly of MNPs. These observations are also in perfect qualitative agreement with the work of Alphand\'ery {\it et al}.~\cite{alphandery2009}. 

\begin{figure}[!htb]
\centering\includegraphics[scale=0.50]{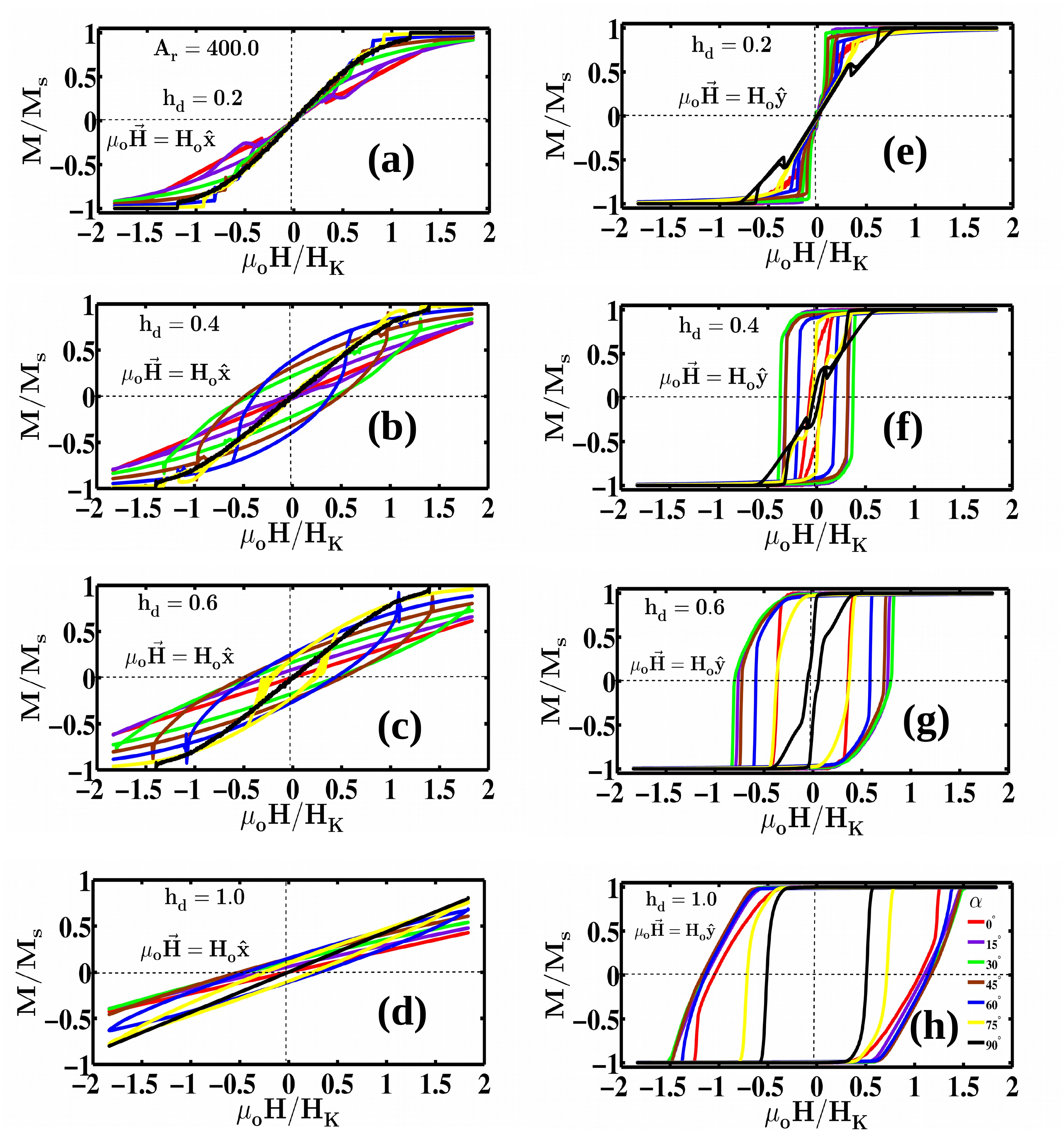}
\caption{Magnetic hysteresis variation with highly anisotropic system, $A^{}_r=400.0$, $h^{}_d$ and external magnetic field direction. We have taken four values of interaction strength: $h^{}_d=0.2$ [(a) and (e)], $h^{}_d=0.4$ [(b) and (f)], $h^{}_d=0.6$ [(c) and (g)], and $h^{}_d=1.0$ [(d) and (h)]. Non-hysteresis is observed with external field along x-direction and most substantial $h^{}_d=1.0$, independent of $\alpha$. Ferromagnetic coupling increases with $h^{}_d$ and external field along the long axis of the system ($y$-direction), irrespective of $\alpha$.} 
\label{figure6}
\end{figure}

The anisotropy axis orientation and external magnetic field direction should strongly affect the hysteresis in the highly anisotropic system. Therefore, we now analyze the hysteresis mechanism in a system with an extremely high value of aspect ratio, i.e. $A^{}_r=400$ in Fig.~(\ref{figure6}). It corresponds to a linear array of dipolar interacting MNPs. In the case of the external magnetic field along the shorter length (along the $x$-axis) and moderate dipolar interaction strength $h^{}_d$, the non-hysteresis is observed with aligned anisotropy ($\alpha=0^\circ$). The hysteresis properties also depend weakly on $\alpha$ and $h^{}_d$ in such a case. Interesting physics emerges with the external magnetic field along the long axis of the system, i.e. $y$-direction. There is an increase in the ferromagnetic interaction with $h^{}_d$, and collinear anisotropy axis ( $\alpha=0^\circ$). Consequently, the hysteresis loop area is exceedingly large, and it increases with $h^{}_d$. The hysteresis loop area also decreases with $\alpha$, but there is a weak dependence on it. The results with external field along the MNPs array axis is in perfect agreement with our recent work~\cite{anand2020}. These are also in excellent qualitative agreement with the work of Serantes {\it et al.}~\cite{serantes2014}. The observation of the anisotropic effect of dipolar interaction with the chain-like array of MNPs is also in agreement with the recent work of Vald\'es {\it et al.}~\cite{valdes2020}.

\begin{figure}[!htb]
	\centering\includegraphics[scale=0.450]{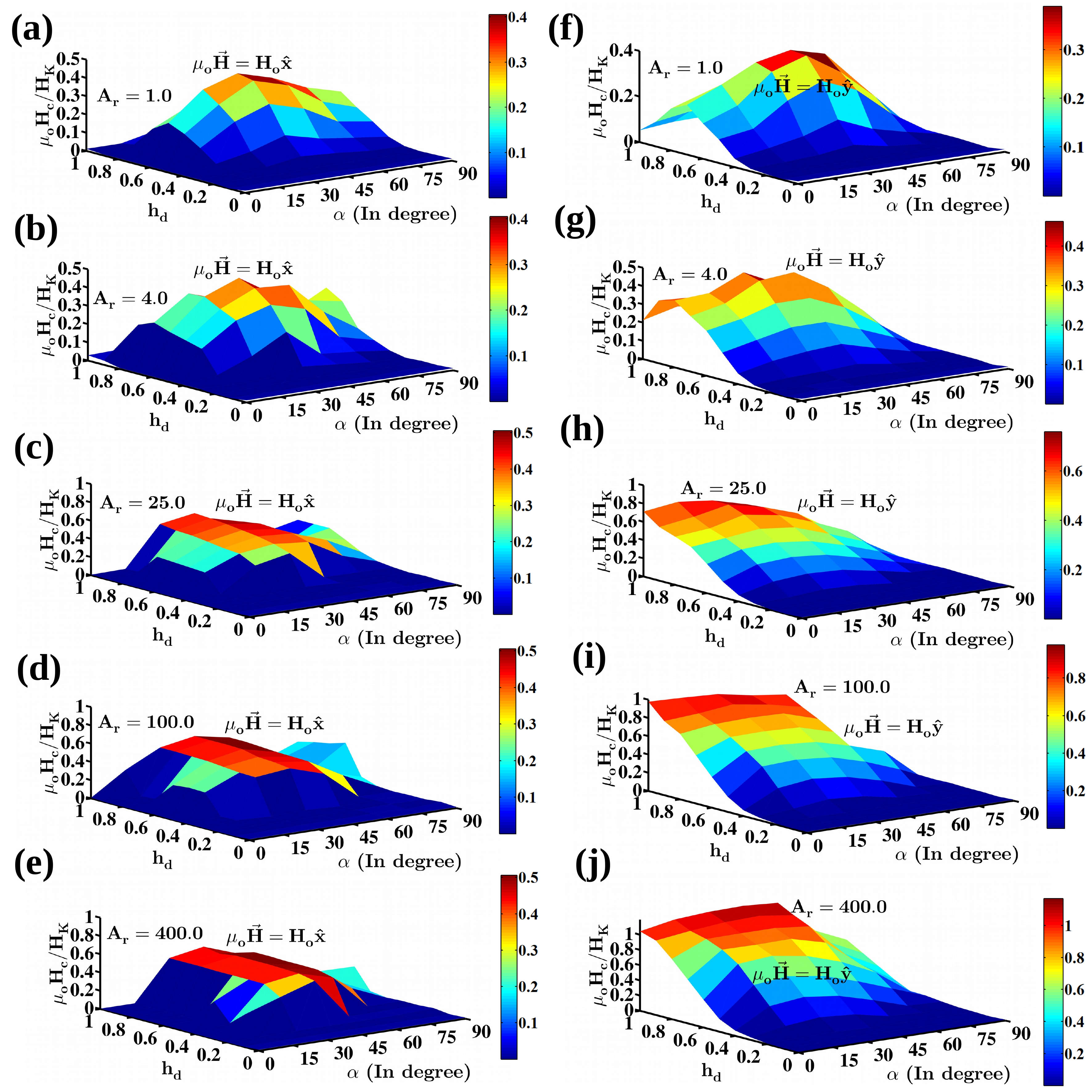}
	\caption{The variation of coercive field $\mu^{}_oH^{}_c$ (scaled by $H^{}_K$) as a function of $h^{}_d$ and $\alpha$ with various $A^{}_r$. The magnetic field is applied along x-axis [(a)-(e)] and y-axis [(f)-(j)]. Irrespective of the applied field direction and weak dipolar interaction ($h^{}_d\leq0.4$), there is an increase in $\mu^{}_oH^{}_c$ with $h^{}_d$ for a fixed $\alpha$, provided the aspect ratio is not very large $A^{}_r\leq4.0$. Notably, $\mu^{}_oH^{}_c$ increases with $h^{}_d$ for a given $\alpha$ and the external field along the $y$-direction.}
	\label{figure7}
\end{figure}

\begin{figure}[!htb]
		\centering\includegraphics[scale=0.450]{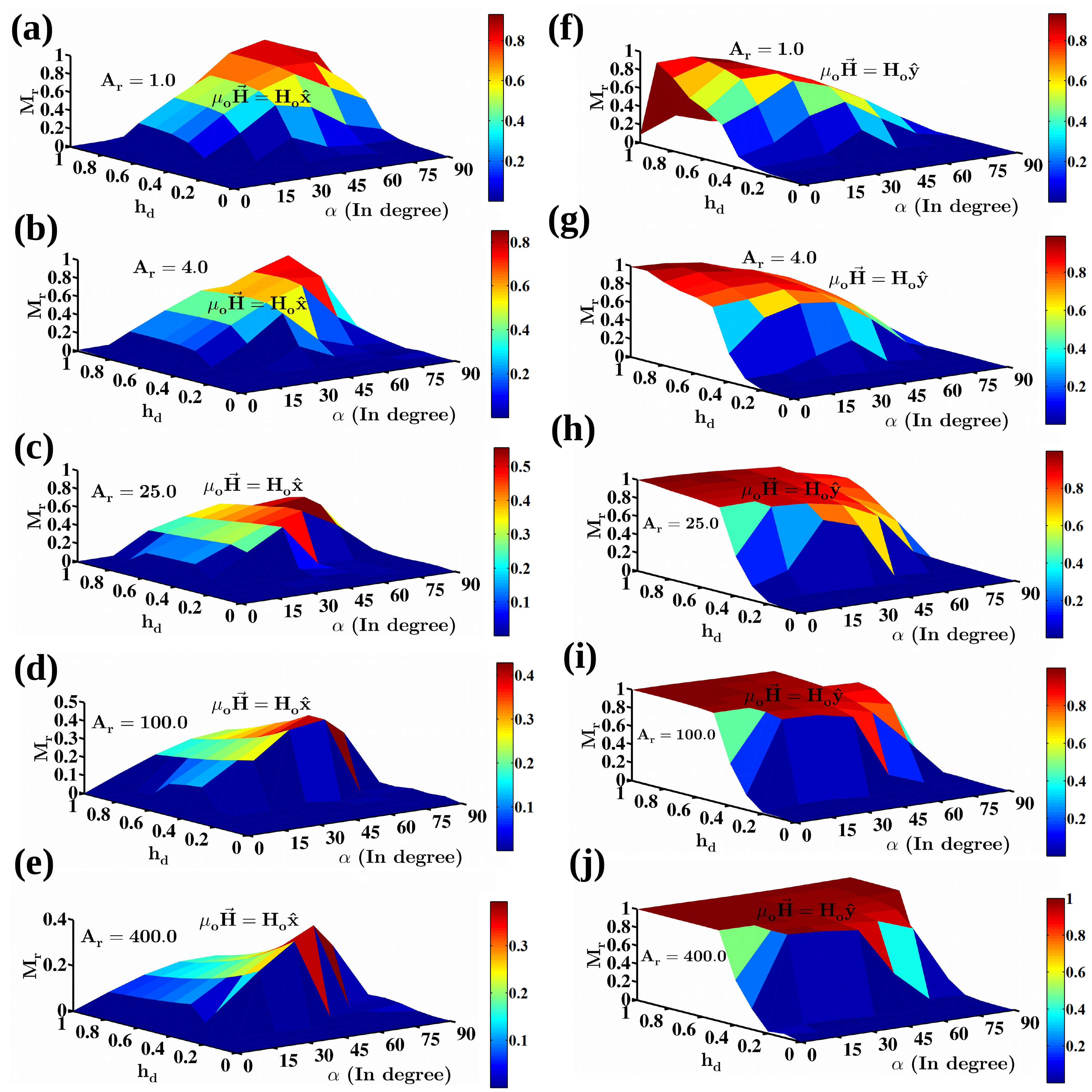}
		\caption{The variation of remanent magnetization $M^{}_r$ as a function of $h^{}_d$ and $\alpha$ with various  $A^{}_r$. The alternating magnetic field is applied along $x$ [(a)-(e)] and $y$ axes [(f)-(j)]. $M^{}_r$ increases with $\alpha$ for a fixed $h^{}_d$ and magnetic field along the $x$-axis, irrespective of $A^{}_r$. There is also an increase in $M^{}_r$ with $h^{}_d$ for large $A^{}_r$ and field applied along the long axis of the system. It is because ferromagnetic coupling enhances with $h^{}_d$ in such a case.} 
		\label{figure8}
	\end{figure}

We now study the coercive field $\mu^{}_oH^{}_c$ variation with  dipolar interaction strength and anisotropy axis orientation to quantify the magnetic hysteresis. Fig.~(\ref{figure7}) shows the  variation of $\mu^{}_oH^{}_c$ as a function of $\alpha$ and $h^{}_d$. We have considered five values of $A^{}_r=1.0$, 4.0, 25.0, 100, and 400. We have also taken into account two cases of magnetic field direction along the $x$ and $y$-axis. The coercive field values are extracted from the corresponding hysteresis curve. In the case of weak dipolar interaction ($h^{}_d\leq0.4$), there is an increase in $\mu^{}_oH^{}_c$ with $h^{}_d$ for a given $\alpha$ and relatively smaller $A^{}_r\leq4.0$, irrespective of magnetic field direction. While for the field applied along the $x$-direction, $\mu^{}_oH^{}_c$ is minimal with weakly interacting MNPs ($h^{}_d\leq0.2$) and large $A^{}_r$, independent of the anisotropy axis orientation $\alpha$. In contrast, it is enormous even with moderate $h^{}_d$ and depends weakly on $\alpha$ and $h^{}_d$ in such a case. Remarkably, $\mu^{}_oH^{}_c$ increases with $h^{}_d$ for a fixed $\alpha$ with the field applied along $y$-direction and significant $A^{}_r$. It is because the ferromagnetic coupling is enhanced with $h^{}_d$. In other words, the dipolar interaction creates an anisotropy termed as shape anisotropy in such a case, resulting in an enhanced coercive field. The study of remanence variation can also provide a better understanding of hysteresis characteristics. Therefore, we now analyze the variation of remanent magnetization $M^{}_r$ as a function of $\alpha$ and $h^{}_d$ in Fig.~(\ref{figure8}). $M^{}_r$ is extremely small with weak dipolar interaction, irrespective of $A^{}_r$ and $\alpha$. In the case of field applied along the $x$-direction, there is an increase in $M^{}_r$ with an increase in anisotropy angle $\alpha$ for a given $h^{}_d$, irrespective of $A^{}_r$. On the other hand, $M^{}_r$ increases with $h^{}_d$ for $A^{}_r=1.0$ and field applied along $y$-direction even with large dipolar interaction strength $h^{}_d\le0.9$, independent of anisotropy axis orientation. There is an increase in $M^{}_r$ with $h^{}_d$ for large $A^{}_r$, irrespective of $\alpha$ because of an enhancement in the ferromagnetic coupling. Remarkably it reaches close to 1.0 with the most substantial dipolar interaction. It means that ferromagnetic coupling strength is the largest in such a case.

\begin{figure}[!htb]
	\centering\includegraphics[scale=0.450]{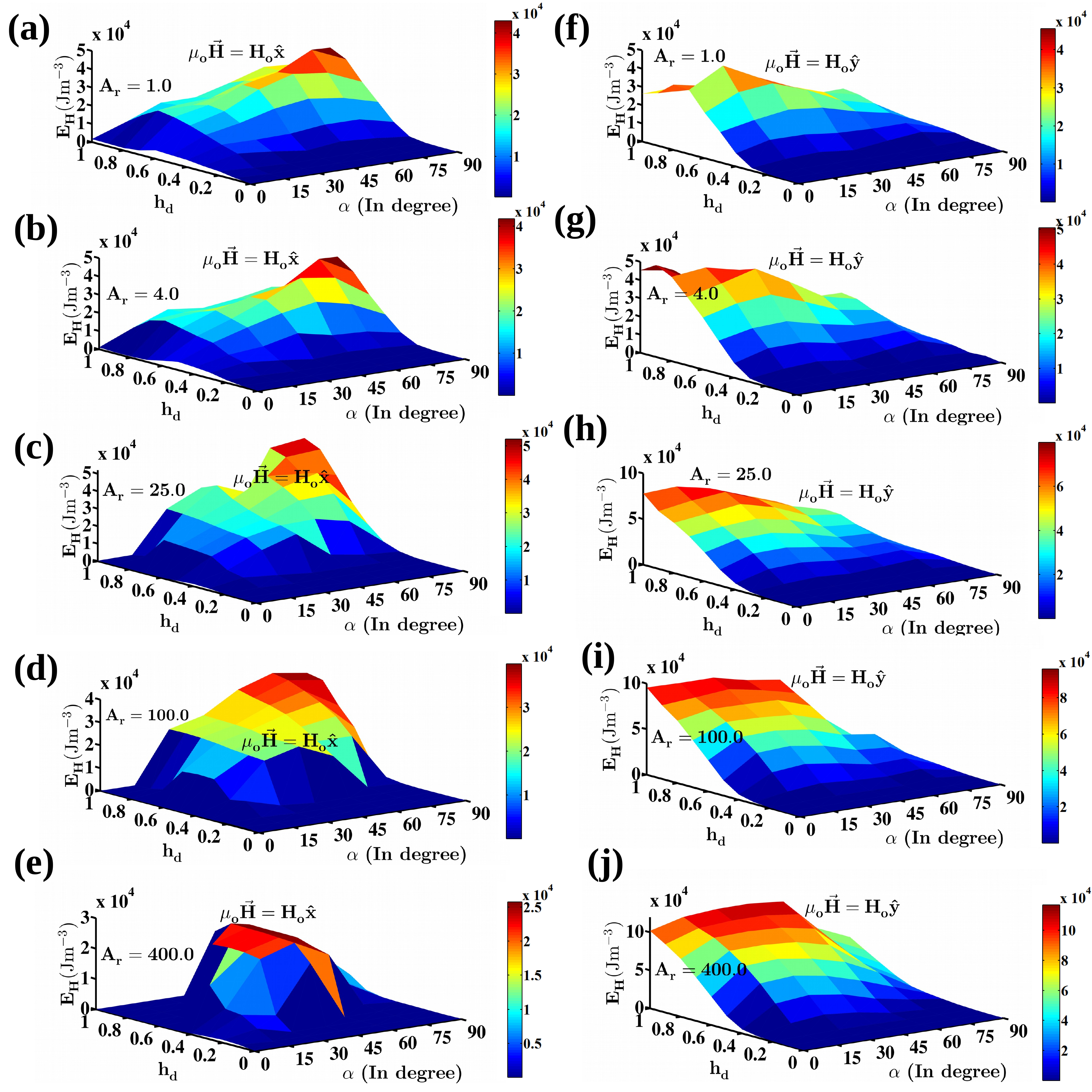}
	\caption{The variation of the amount of heat dissipated $E^{}_H$ with $h^{}_d$ and $\alpha$ for various $A^{}_r$. The external oscillating magnetic field is applied along x-axis [(a)-(e)] and y-axis [(f)-(j)].  $E^{}_H$ is minimal for negligible and weak dipolar interaction, independent of $A^{}_r$, $\alpha$, and external field direction. There is an increase in $E^{}_H$ with $h^{}_d$ and large $A^{}_r$, provided the alternating magnetic field is along the long axis of the system, irrespective of $\alpha$.} 
	\label{figure9}
\end{figure}

Finally, we study the amount of heat dissipation $E^{}_H$ variation with $h^{}_d$ and $\alpha$ in Fig.~(\ref{figure9}). The value of $E^{}_H$ is minimal for negligible and weak magnetic interaction strength. In the case of the field applied along the $x$-axis, $E^{}_H$ increases with $\alpha$ for a moderate and fixed $h^{}_d$. While for the field applied along the $y$-direction, $E^{}_H$ increases with $h^{}_d$ for a fixed $\alpha$ and relatively smaller $A^{}_r\leq4.0$, provided $h_d\leq0.8$. It starts to diminish as $h^{}_d$ is further increased, indicating the dominance of antiferromagnetic coupling. Remarkably, the value of $E^{}_H$ increases with $h^{}_d$ and large $A^{}_r$, irrespective of $\alpha$. It is because the dipolar interaction promotes ferromagnetic coupling in such a system. Notably, $E^{}_H$ is exceedingly large with the highly anisotropic system $A^{}_r=400$. The decrease of $E^{}_H$ with $\alpha$ for the external magnetic field along the $y$-axis is in perfect qualitative agreement with the work of Serantes {\it et al.}~\cite{serantes2014}. It is also in perfect agreement with our recent work~\cite{anand2020}. Conde-Lebor\'an {\it et al.} also obtained similar results  in a linear array of MNPs ~\cite{conde2015}. Our results also agree well with them.

\section{Summary and conclusion}
We have investigated magnetic hysteresis in the two-dimensional ($l^{}_x\times l^{}_y$) array of nanoparticles using numerical simulations with aligned anisotropy. In particular, we probed the hysteresis mechanism as a function of aspect ratio $A^{}_r=l^{}_y/l^{}_x$, anisotropy axis orientational angle $\alpha$, magnetic interaction strength $h^{}_d$, and the applied alternating magnetic field direction using kinetic Monte Carlo simulations. The assumption of an aligned anisotropy axis provides an elegant way to investigate the precise role of anisotropy and dipolar interaction on the hysteresis characteristics in such an ordered assembly of MNPs.
The superparamagnetic character is dominant with small dipolar interaction ($h^{}_d\leq0.2$), resulting in the minimal hysteresis loop area. These observations are also robust to the aspect ratio, anisotropy axis orientation, and the external magnetic field directions. Interestingly, the hysteresis behaviour with the external field along the $x$-direction and relatively smaller $A^{}_r$ is precisely the same as with the field along the $y$-direction for complementary orientational angle $\alpha$. It implies that the hysteresis characteristics with $\alpha=0^{\circ}$ and the field applied along the $x$-direction are the same as that of $\alpha=90$ and external magnetic field along the $y$-direction. One can draw similar conclusions for other complementary angles also. Remarkably, double loop hysteresis emerges even with moderate dipolar interaction strength $h^{}_d\approx0.4$, reminiscent of antiferromagnetic coupling. These features are strongly dependent on the $\alpha$ and aspect ratio of the system. The dominance of the antiferromagnetic coupling is the strongest with the most substantial dipolar interaction strength ($h^{}_d=1.0$), irrespective of $\alpha$ and magnetic field direction. Interestingly, non-hysteresis is observed in the highly anisotropic system ($A^{}_r=400$) even with small dipolar interaction strength and external magnetic along the shorter axis of the sample ($x$-direction). On the other hand, the dipolar interaction promotes the ferromagnetic coupling with the applied field along the $y$-direction, the long axis of the sample in such a system. Consequently, the hysteresis loop area increases with $h^{}_d$, and it has a weak dependence on the anisotropy axis orientation. Ordered arrays of MNPs with aligned anisotropy axes are frequently observed in experiments~\cite{amali2011,hu2015,jain2016}. We believe that our results could be beneficial in predicting the hysteresis properties of such MNPs ensembles.

The study of coercive field $\mu^{}_oH^{}_c$, remanent magnetization $M^{}_r$, and the heat dissipation $E^{}_H$ also provides vital quantitative information regarding hysteresis. Irrespective of the applied field direction and weak dipolar interaction ($h^{}_d\leq0.4$), there is an increase in $\mu^{}_oH^{}_c$ with $h^{}_d$ for a fixed $\alpha$, provided the aspect ratio is not very large $A^{}_r\leq4.0$. Notably, $\mu^{}_oH^{}_c$ increases with $h^{}_d$ for a given $\alpha$ and the external field along the $y$-direction. In such a case, the dipolar interaction creates an additional anisotropy, known as the shape anisotropy, resulting in an enhancement in the coercive field. $M^{}_r$ increases with $\alpha$ for a fixed $h^{}_d$ and magnetic field along the $x$-axis, irrespective of $A^{}_r$. There is also an increase in $M^{}_r$ with $h^{}_d$ for large $A^{}_r$ and field applied along the long axis of the system. It is because ferromagnetic coupling enhances with $h^{}_d$ in such a case. The amount of heat dissipation $E^{}_H$ due to the hysteresis also strongly depends on these parameters. $E^{}_H$ is minimal for negligible and weak dipolar interaction, irrespective of $A^{}_r$, $\alpha$, and external field direction. Conversely, $E^{}_H$ increases with $\alpha$ and $h^{}_d$ ($h^{}_d\leq0.4$), provided the external field is along the $x$-axis. Remarkably, there is an enhancement in $E^{}_H$ with $h^{}_d$ and large $A^{}_r$ and magnetic field applied along the long axis of the system, irrespective of $\alpha$. In such a case, the dipolar interaction induces ferromagnetic coupling, resulting in increased heat dissipation. 

In  conclusion, we analyzed the hysteresis characteristics in the ordered arrays of nanoparticles with aligned anisotropy axes using extensive numerical simulations. We have also studied the hysteresis response as a function of dipolar interaction strength, the system's aspect ratio, and the external magnetic field directions. The hysteresis properties depend strongly on these parameters. The assumption of common easy axes is one of the elegant routes to probe the precise role of anisotropy and dipolar interactions. Such systems are also realized in experiments and beneficial in diverse technological applications~\cite{fragouli2010,lisjak2018}. We believe that our results are extremely useful in these contexts. {\color{black}Therefore, we believe that the present work could instigate combined efforts in experimental, analytical, and computational research for these precious and physics enriched systems.}
\section*{ACKNOWLEDGMENTS}
Most of the numerical simulations presented in this work have been performed in the Department of Physics, Indian Institute of Technology (IIT) Delhi, India. The author is thankful to Prof. Varsha Banerjee for providing the computational facilities at IIT Delhi.
\section*{DATA AVAILABILITY}
The data that support the findings of this study are available from the corresponding author upon reasonable request.
\bibliography{ref}
\newpage
\newpage

\newpage

\newpage

\newpage
\newpage

\newpage
	
	\newpage

\newpage
\end{document}